\documentclass[sigconf]{acmart}

\usepackage{comment}
\usepackage{caption}
\usepackage{subcaption}
\usepackage{amsfonts}
\usepackage{multirow}
\usepackage{enumerate}
\usepackage{rotating}

\AtBeginDocument{%
  \providecommand\BibTeX{{%
    \normalfont B\kern-0.5em{\scshape i\kern-0.25em b}\kern-0.8em\TeX}}}

\setcopyright{acmcopyright}
\copyrightyear{2020}
\acmYear{2020}
\acmDOI{XXXX}

\acmConference[MMArt-ACM '20] {2020 Joint Workshop on Multimedia Artworks Analysis and Attractiveness Computing in Multimedia}{June 8, 2020}{Dublin, Ireland}
\acmBooktitle{2020 Joint Workshop on Multimedia Artworks Analysis and Attractiveness Computing in Multimedia (MMArt-ACM'20), June 8, 2020, Dublin, Ireland}

\begin{document}
\fancyhead{}
\title{Iconify: Converting Photographs into Icons}

\author{Takuro Karamatsu}
\affiliation{%
  \institution{Kyushu University}
  \city{Fukuoka}
   \country{Japan}
}

\author{Gibran Benitez-Garcia}
\author{Keiji Yanai}
\affiliation{%
  \institution{The University of Electro-Communications}
  \city{Tokyo}
  \country{Japan}}
\email{yanai@cs.uec.ac.jp}

\author{Seiichi Uchida}
\affiliation{%
  \institution{Kyushu University}
  \city{Fukuoka}
  \country{Japan}
}
\email{uchida@ait.kyushu-u.ac.jp}

\renewcommand{\shortauthors}{Karamatsu, et al.}

\begin{abstract}
In this paper, we tackle a challenging domain conversion task between photo and icon images. 
Although icons often originate from real object images (i.e., photographs), severe abstractions and simplifications are applied to generate icon images by professional graphic designers.
Moreover, there is no one-to-one correspondence between the two domains, for this reason we cannot use it as the ground-truth for learning a direct conversion function. 
Since generative adversarial networks (GAN) can undertake the problem of domain conversion without any correspondence, we test CycleGAN and UNIT to generate icons from objects segmented from photo images.
Our experiments with several image datasets prove that CycleGAN learns sufficient abstraction and simplification ability to generate icon-like images. 
\end{abstract}


\begin{teaserfigure}
\centering
  \includegraphics[width=0.95\textwidth]{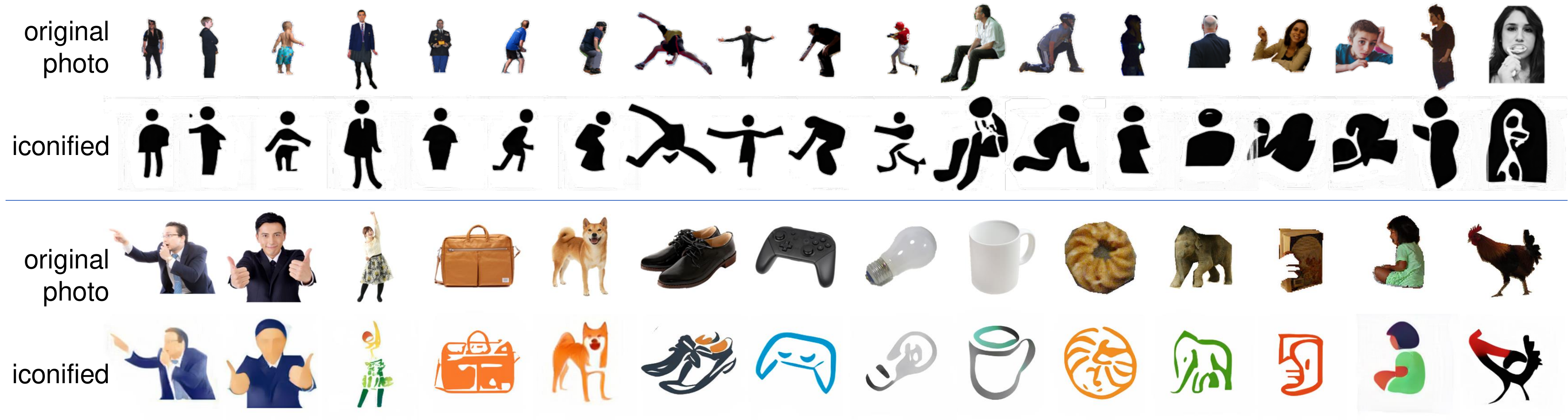}\\[-3mm]
  \caption{Iconified results. Black-and-white icons (top) and color icons (bottom) are generated from photos.}
  \Description{Machine learning techniques can capture
and mimic the abstraction and simplification skill of human experts on designing icons.}
  \label{fig:teaser}
\end{teaserfigure}

\maketitle
\section{Introduction\label{sec:intro}}
\begin{figure}[tb] 
    \centering
    \includegraphics[width=0.4\textwidth]{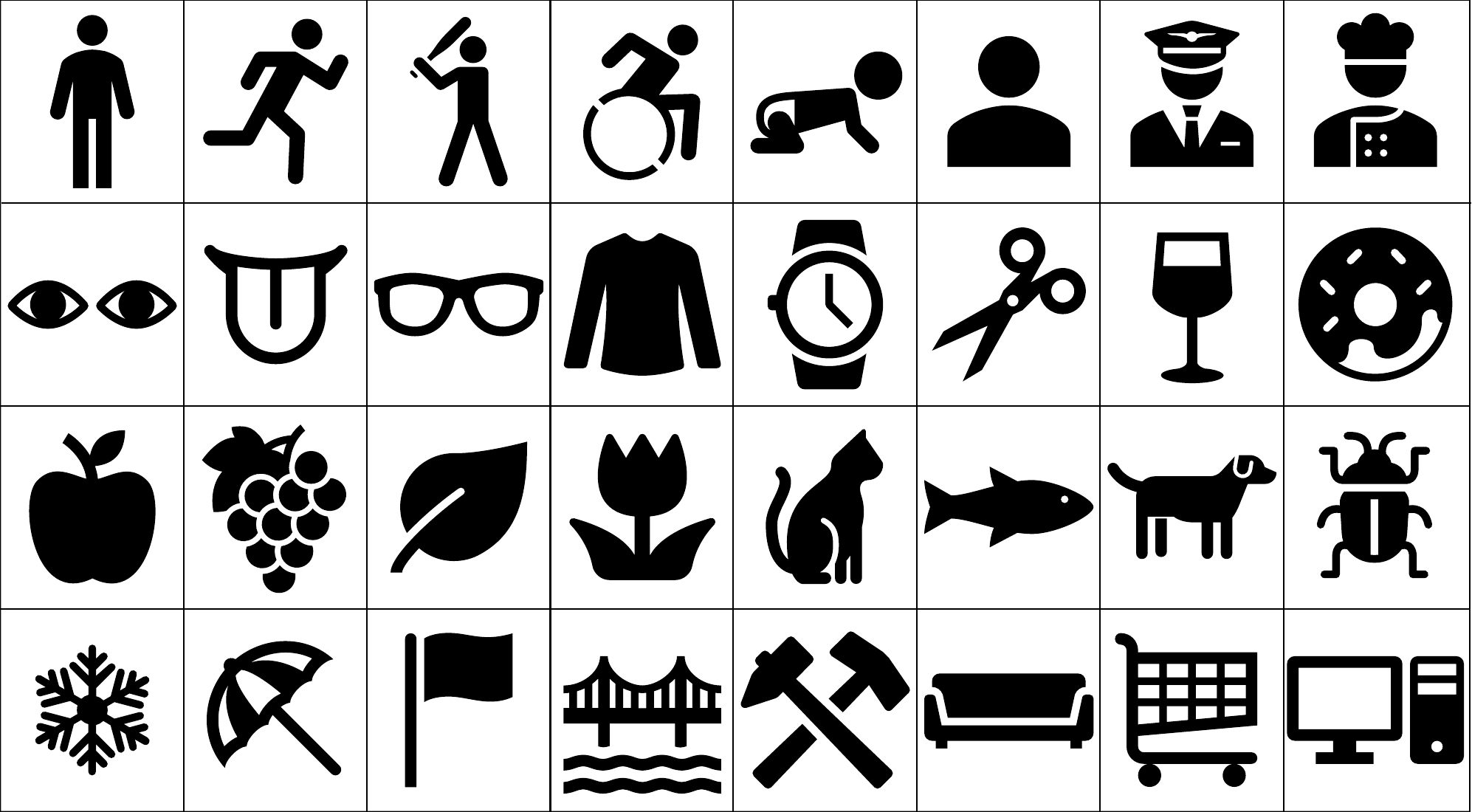}\\[-2mm]
    \caption{Black-and-white icon images provided in Microsoft PowerPoint.}
    \label{fig:icon_sample}
\end{figure}

Throughout this paper, we assume that icon images, or pictogram, are designed by abstracting and simplifying some object images. 
Figure~\ref{fig:icon_sample} shows the black-and-white icon images provided in Microsoft PowerPoint. We can observe that icon images are not just binarized object images but designed with severe abstraction and simplification of the original object appearance. For example, person's heads are often drawn as a plain circle. Graphic designers have professional knowledge and skills of abstraction and simplification while keeping discriminability as the original object.
\par

This paper reports our trials to generate icon images automatically from natural photographs by using machine learning techniques. 
Our main purpose is to reveal whether the machine learning techniques can capture and mimic the abstraction and simplification skill of human experts on designing icons. 
We encounter the following three difficulties that make our task challenging.
\par
The first difficulty is that this is a domain conversion task between two sample sets (i.e., domains). 
If we have a dataset with image pairs of an icon and its original photo image, our image generation task becomes a direct conversion, 
which can be solved by conventional methods, such as U-net or its versions. 
However, it is not feasible to have such a dataset in practice. 
Hence, we only can prepare a set of photo images and a set of icon images, without any one-to-one correspondence between the two domains.
\par
The second difficulty lies in the large style difference between the photo image domain and the icon image domain.
For example, the appearance of a person's head is totally different than that represented in icon images, as shown in Figure~\ref{fig:icon_sample}.
Thus, the selected machine learning technique must be able to learn a mapping to fill the large gap between both domains.
\par
The third difficulty lies in the large appearance variations in both domains. 
Although icon images are simple and plain, they still have large variations in their shapes to represent various objects. 
Object photo images have even more variations in their shape, color, texture, etc. 
The mapping between the two domains needs to cope with these variations.
\par
We, therefore, employ CycleGAN\cite{CycleGAN} and UNIT\cite{UNIT} as the machine learning techniques for our task. 
Both of them can learn the mapping between the two different domains thanks to a cycle-consistency loss, and this mapping can be used as a domain converter. 
Note that the original papers of CycleGAN and UNIT tackle rather easier domain conversion tasks, such as horse and zebra and winter and summer scenery.
On the other hand, for our task, they have to learn the mapping between a photo image set and an icon image set.
So that, the learned mapping can convert arbitrary objects from the photo image to its iconified version.
\par
The results of our trials with several image datasets reveal that CycleGAN is able to iconify photo images even with the mentioned difficulties, as shown in Figure~\ref{fig:teaser}. 
This proves that CycleGAN can lean the abstraction and simplification ability. 
We also reveal that the quality of the generated icons can be improved by limiting both domains to a specific object, such as persons.
\section{Related work\label{sec:related}}
\subsection{Logos and icons}
To the best of our knowledge, there is no computer science research for icons generation, which are defined as abstracted and simplified object images. 
Instead, we can find many research trials about {\em logo}.  In \cite{logo_definition}, logo is defined as ``a symbol, a graphic and visual sign which plays an important role into the communication structure of a company'' and classified into three types: Iconic or symbolic logo, text-based logo, and mixed logo. In this sense, logo is a broader target than icon for visual analytics research. \par
Comparing to traditional logo design researches that often focus how the logo design affects human behavior and impression through subjective experiments (e.g., \cite{logo_development,logo_evaluation,logo_move,logo_change}), recent researches become more objective and data-driven.
Those works are supported by different logo image datasets, such as FlickrLogos\cite{FlickrLogos}, LOGO-net\cite{LOGO-net}, WebLogo-2M\cite{WebLogo-2M}, Logo-2K+\cite{Logo-2K+}, and LLD\cite{LLD}. 
Especially, LLD is comprised of 6 million logo images and sufficient as a dataset for data-hungry machine learning techniques.
 
\subsection{Image generation by machine learning}
After the proposal of variational autoencoder (VAE), Neural Style Transfer (NST)~\cite{styletransfer} and generative adversarial networks\linebreak (GAN), many image generation methods based on machine learning have been proposed. Especially, GAN-based image generation is a big research trend, while being supported by many quality improvement technologies, such as \cite{WGAN,PGGAN,SinGAN}. 
\par
GANs are also extended to deal with image conversion tasks. 
Pix2pix~\cite{pix2pix} is a well-known technique for converting an input image from a domain $X$ to an image in a domain $Y$. Pix2pix is trained with a ``paired'' sample set $\{(x,y)\| x\in X, y\in Y\}$. For example, $x$ is a scene image during daytime and $y$ is a nighttime image at the same location. 
By training pix2pix with such pairs, a day-night converter can be performed.
CycleGAN\cite{CycleGAN} and UNIT\cite{UNIT} can also realize a domain conversion task but they are more advanced than pix2pix. 
Just given two sample sets (i.e., two domains) and without any correspondence between them, they can learn a mapping function between both domains.
\par
Those image generation and conversion methods are also used for generating visual designs. For example, the idea of NST is applied to attach decoration to font images~\cite{fontST} and logo skeleton~\cite{tugs}. GAN is applied to font generation~\cite{fontGAN,hayashi}. In \cite{icon_color}, 
a conditional GAN is proposed to paint an edge image with  
a similar color style to a color image. In \cite{LLD}, GANs are used to generate general logo images from random vectors.
In \cite{muhammad}, reinforcement learning is employed for sketch abstraction.
\par
In this paper, we treat an icon generation task as a domain conversion between the photo image domain and the icon image domain. 
Since there is no prior correspondence between them, we employ CycleGAN~\cite{CycleGAN} and UNIT~\cite{UNIT}. 
We will see that those GANs can bridge the huge gap between the two domains and establish a mapping that ``iconify'' a photo image to an icon-like image.

\section{GANs to Iconify\label{sec:GAN}}
We employ CycleGAN\cite{CycleGAN} and UNIT\cite{UNIT} to transform natural photos to icon-like images. Both of them are a domain conversion method and can determine a mapping between two domains (i.e., image sets) without giving one-to-one correspondence between the elements of the two sets. 
In our task, it is not feasible to give one-to-one correspondence between a photo and an icon image in advance to training. Therefore CycleGAN and UNIT are reasonable choices.

\subsection{CycleGAN}
\begin{figure}[t] 
    \centering
    \includegraphics[width=0.4\textwidth]{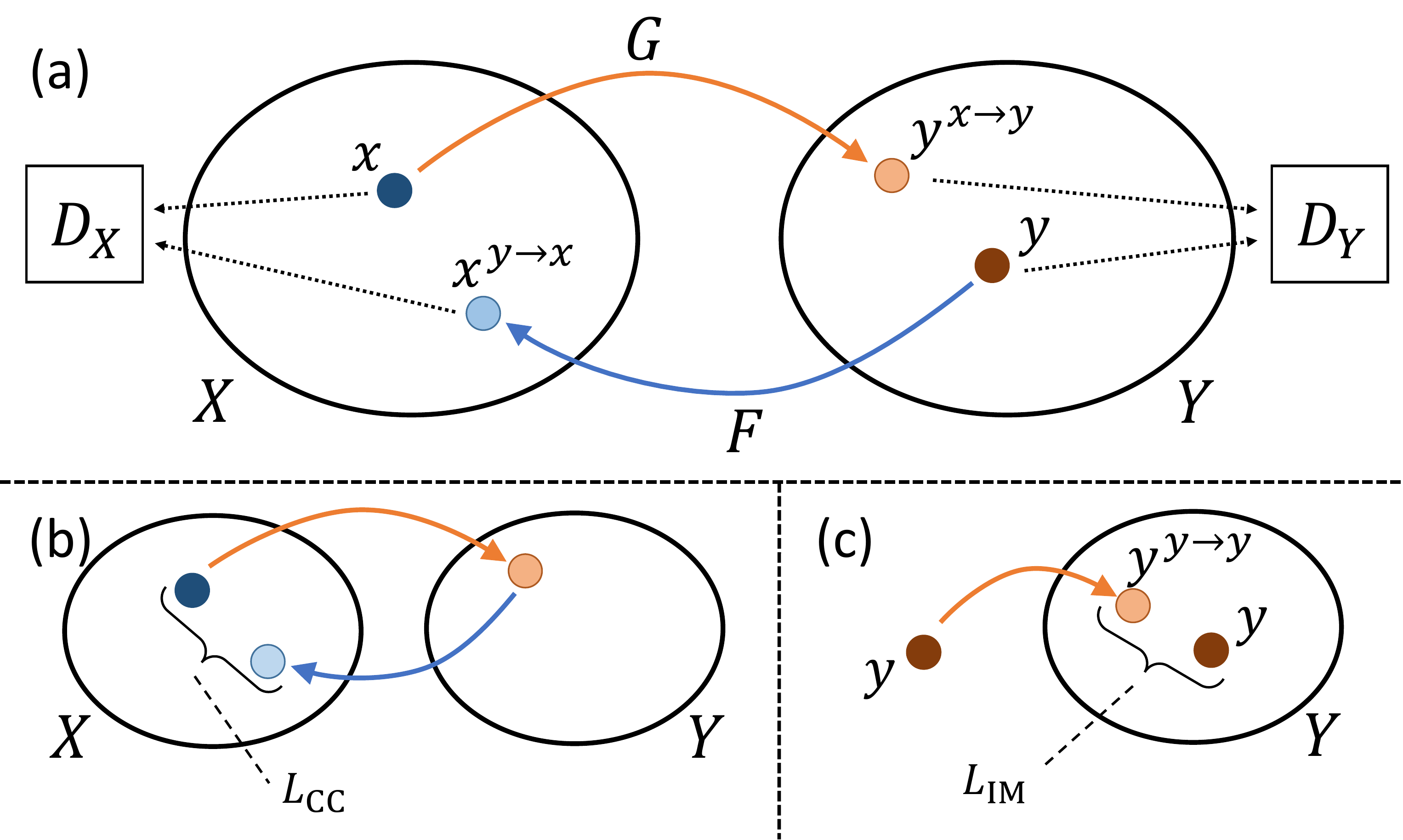}
    \caption{(a)~Overview of CycleGAN~\cite{CycleGAN}. Two GANs are coupled to bridge two domains $X$ and $Y$. (b)~Cycle-consistency loss, $L_{\mathrm CC}$. (c)~Identity mapping loss, $L_{\mathrm IM}$.}
    \label{fig:CycleGAN}
\end{figure}

CycleGAN\cite{CycleGAN} determines a mapping between two image sets, $X$ and $Y$, without giving any image-to-image correspondence. 
Figure~\ref{fig:CycleGAN} illustrates the overall structure of CycleGAN, 
which is comprised of two generators (i.e., style transformers) $G$ and $F$ and two iscriminators $D_X$ and $D_Y$. 
In other words, two GANs ($G\leftrightarrow D_Y$ and $F\leftrightarrow D_X$) are coupled to bridge two domains $X$ and $Y$. 
\par
Those modules are co-trained by three loss functions: the adversarial loss $L_{\mathrm GAN}$, the cycle-consistency loss $L_{\mathrm CC}$, and the identity mapping loss $L_{\mathrm IM}$.
The adversarial loss is used for training two GANs.
The cycle-consistency loss is necessary to realize a bi-directional and one-to-one mapping between $X$ and $Y$ by letting $G^{-1}\sim F$ and vice versa.
The identity mapping loss is an optional loss and used 
for the color constancy on the style transformation by $F$ and $G$. \par
In the following experiment, we use the network structure and the original implementation\footnote{https://github.com/junyanz/pytorch-CycleGAN-and-pix2pix} provided by the authors \cite{CycleGAN}. 
Note that for the experiments to generate black-and-white icons from color photos (Sections \ref{sec:ex1} and \ref{sec:ex2}), the color constancy is not necessary. Therefore we weaken the identity mapping loss for those experiments.

\subsection{UNIT}
\begin{figure}[tb] 
    \centering
    \includegraphics[width=1\linewidth]{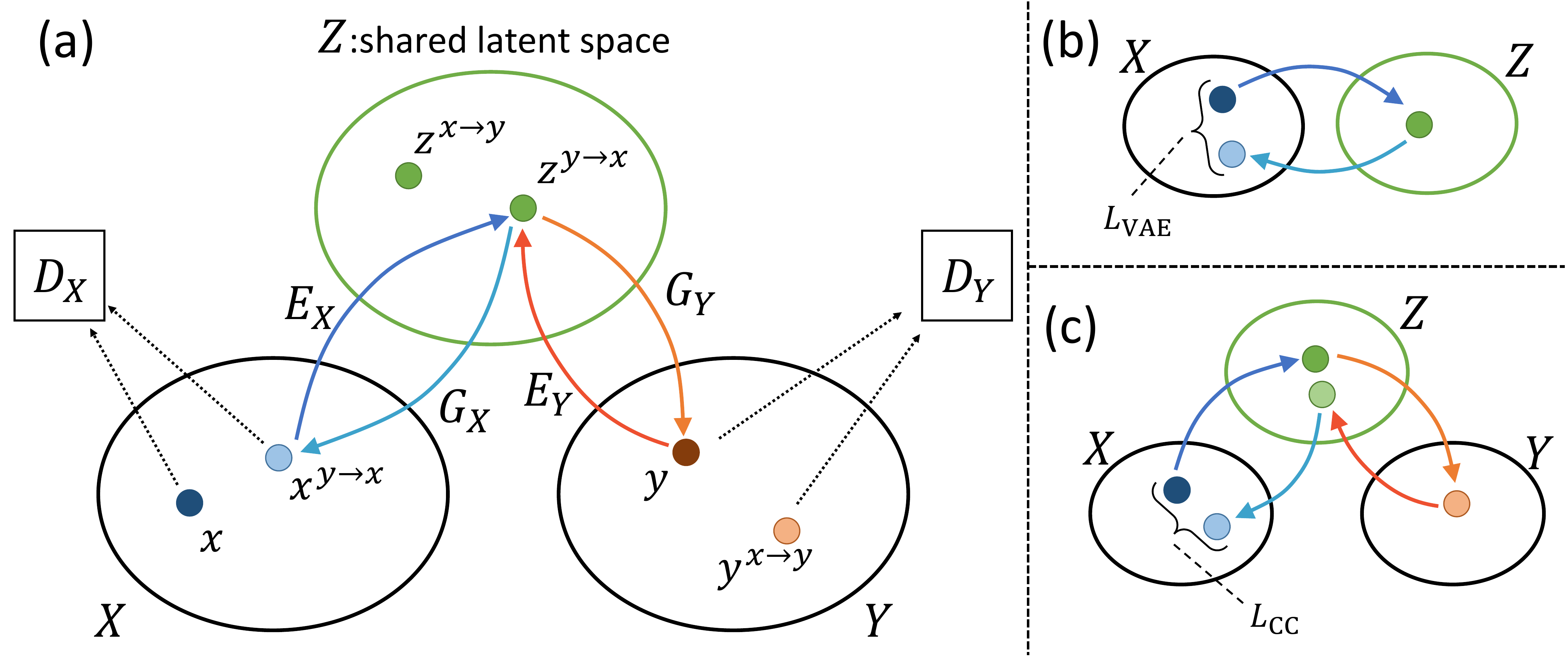}
    \caption{(a)~Overview of UNIT~\cite{UNIT}. (b)~VAE-loss, $L_{\mathrm{VAE}}$. (c)~Cycle-Consistency loss, $L_{\mathrm{CC}}$.}
    \label{fig:UNIT}
\end{figure}

UNIT~\cite{UNIT} can be considered as an extended version of CycleGAN, which accomplish style transformation between two image sets, $X$ and $Y$.
Its main difference from CycleGAN is the condition that an original image and its transformed image should be represented by the same variable in the latent space $Z$.
As illustrated in Figure~\ref{fig:UNIT}, UNIT is comprised of two encoders $E_X$ and $E_Y$, two generators $G_X$ and $G_Y$ and two discriminators $D_X$ and $D_Y$. 
Note that the generator $G$ of CycleGAN is divided into $E_X$ and $G_Y$ in UNIT. 
Those modules are co-trained by VAE loss $L_{\mathrm VAE}$, adversarial loss $L_{\mathrm GAN}$, and cycle-consistency loss $L_{\mathrm CC}$.
The VAE loss is introduced so that the latent variable contains sufficient information of original images. 
In the following experiment, we use the network structure and the original implementation\footnote{https://github.com/mingyuliutw/UNIT} provided by the authors \cite{UNIT}.

\section{Image Datasets to Iconify}
\subsection{Object photograph data\label{sec:obj_sample}}
\begin{figure}[tb] 
    \centering
    \includegraphics[width=\linewidth]{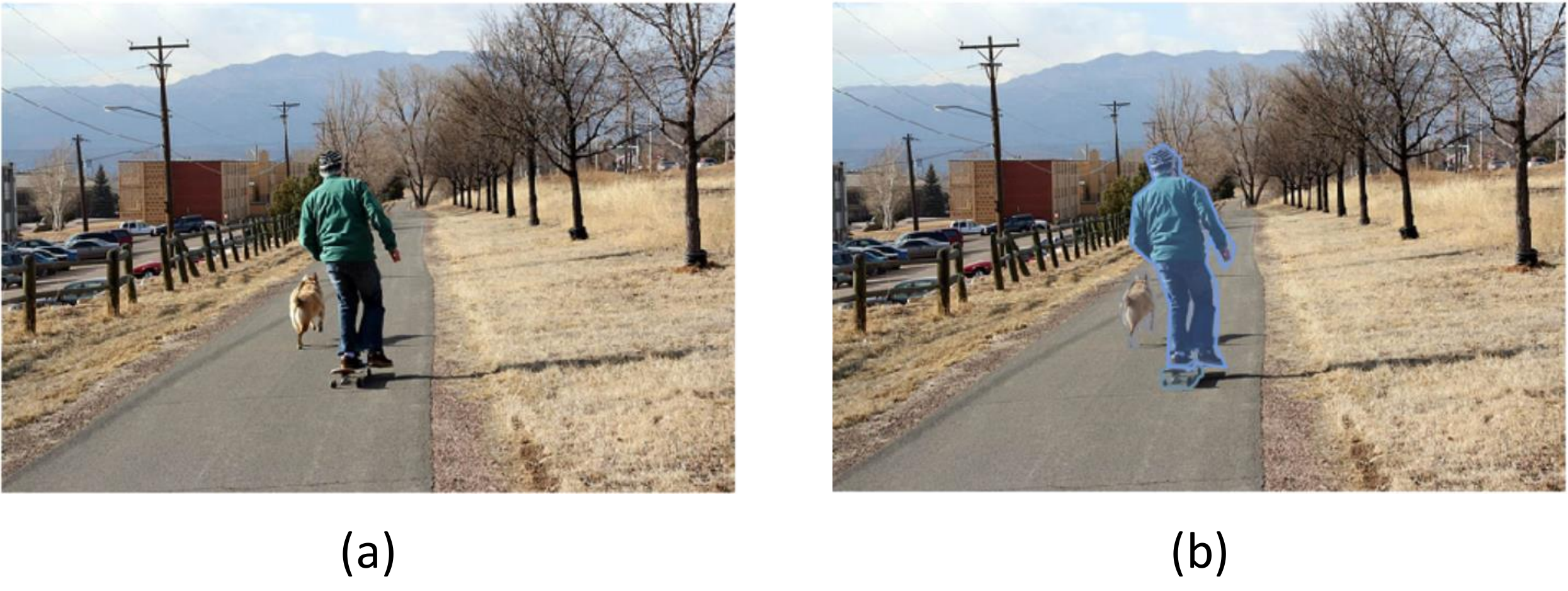}
    \\[-5mm]
    \caption{(a)~An image from MS-COCO. Three object class labels, ``person'', ``dog'', and ``skateboard'', are attached to this image. (b)~Pixel-level ground-truth for those three classes.}
    \label{fig:COCO_sample}
\bigskip\bigskip        
    \includegraphics[width=0.9\linewidth]{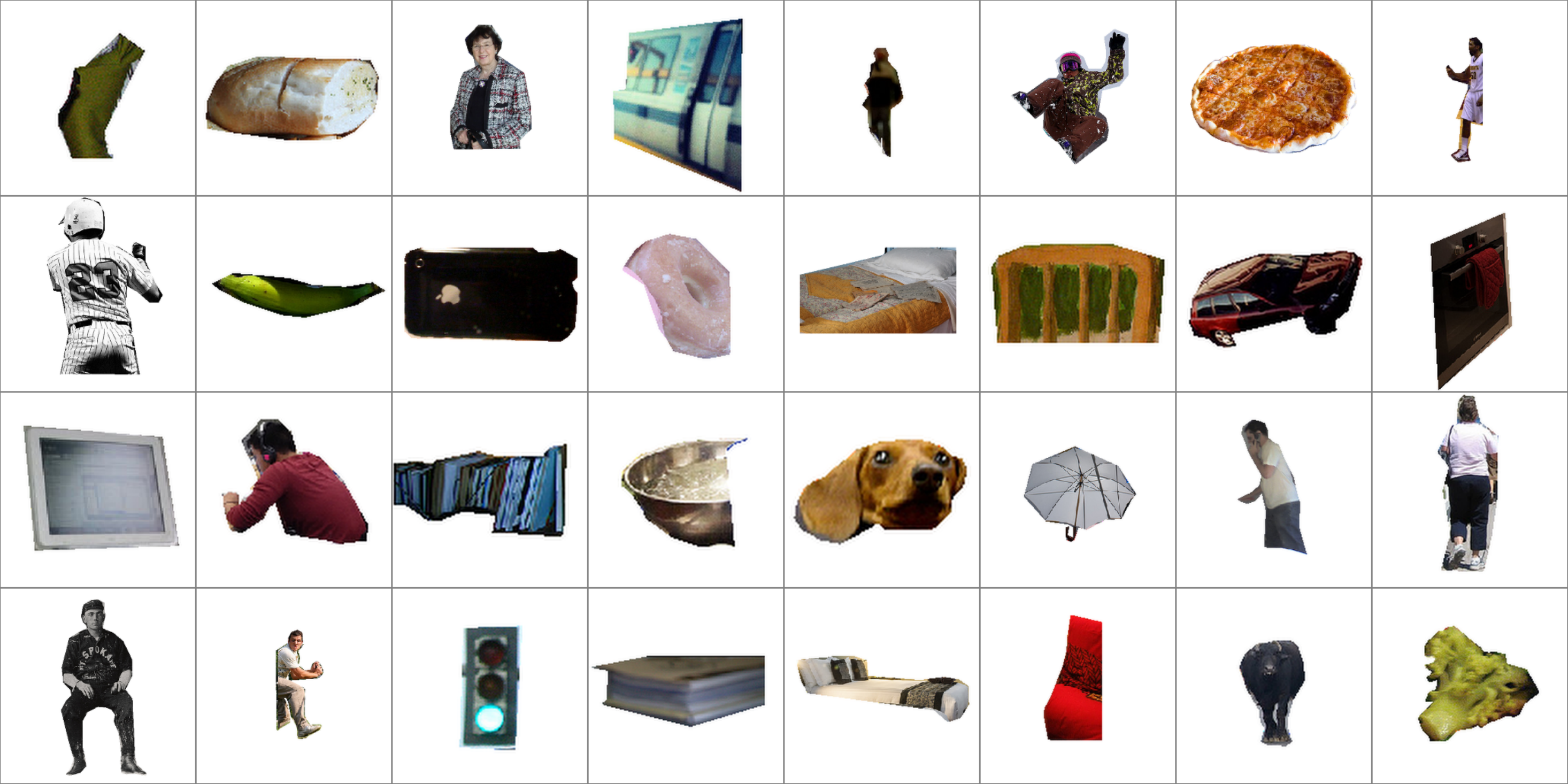}\\[-3mm]
    \caption{Object photo images extracted from MS-COCO.}
    \label{fig:obj_sample}
\bigskip\bigskip
    \includegraphics[width=0.8\linewidth]{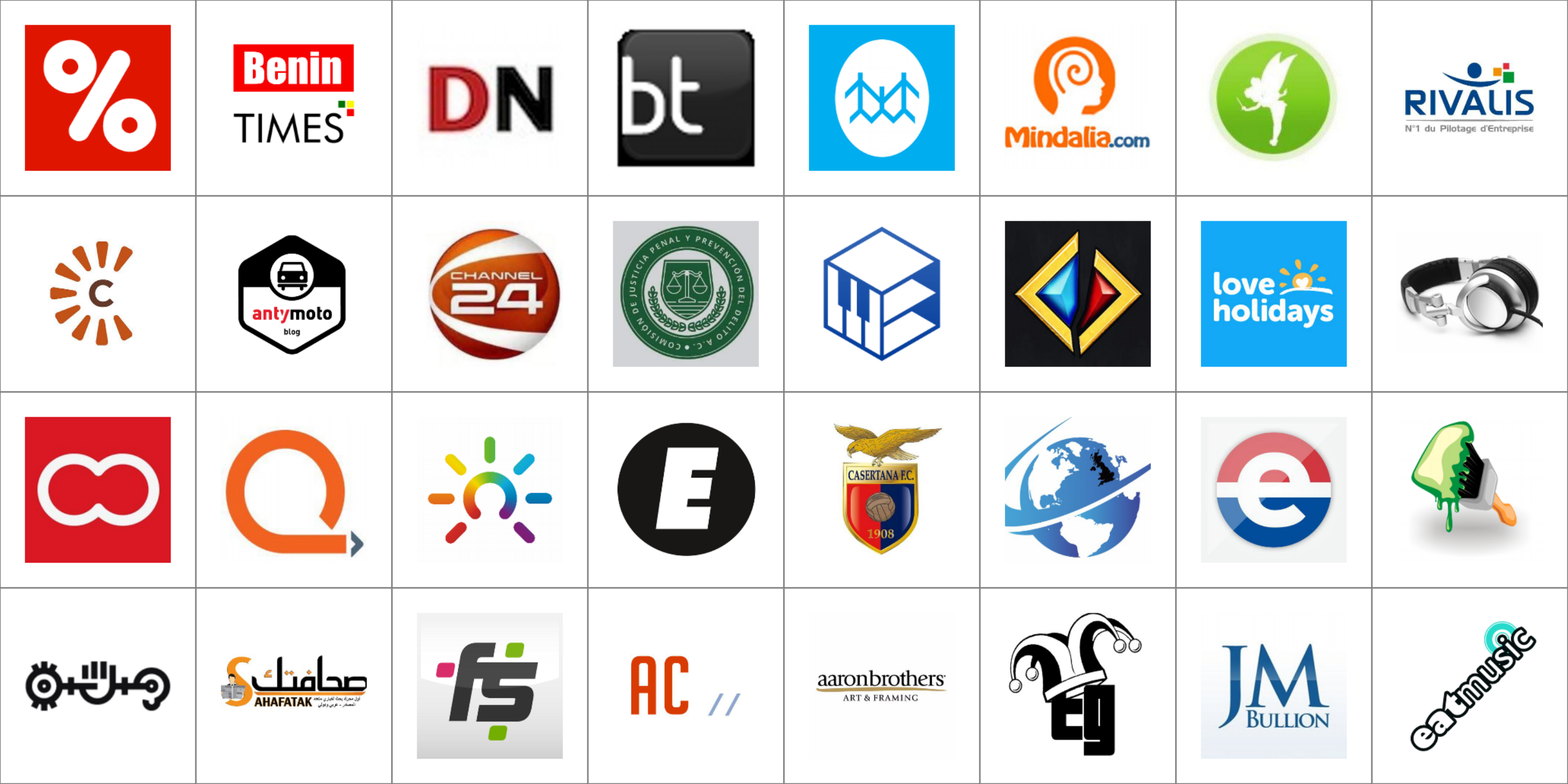}\\[-3mm]
    \caption{Logo images from LLD~\cite{LLD}.}
    \label{fig:logo_sample}
\bigskip\bigskip
  \begin{minipage}{0.23\textwidth}
      \centering
        \includegraphics[width=\textwidth]{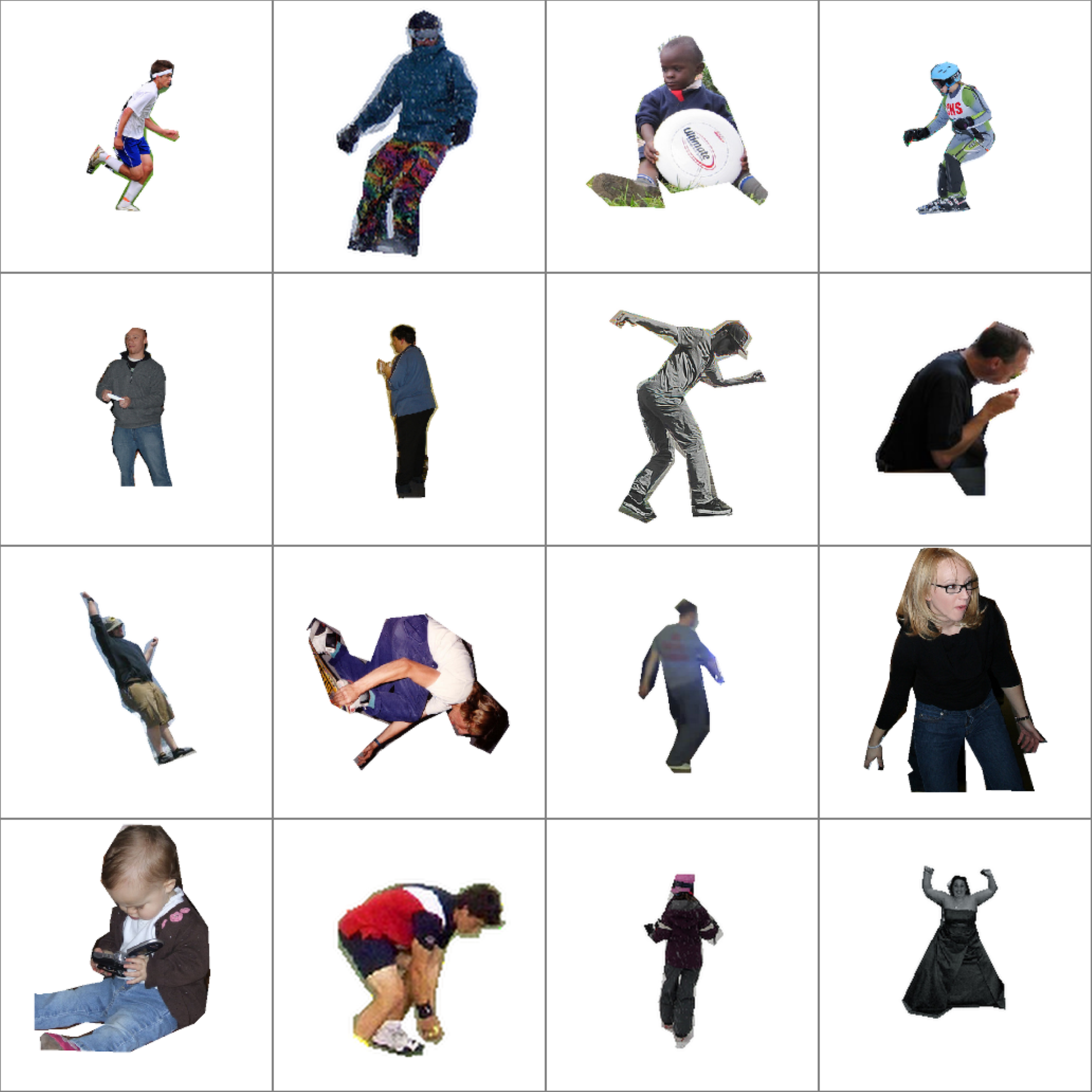}
  \small{(a)}
\end{minipage}
\begin{minipage}{0.23\textwidth}
      \centering   
  \includegraphics[width=\textwidth]{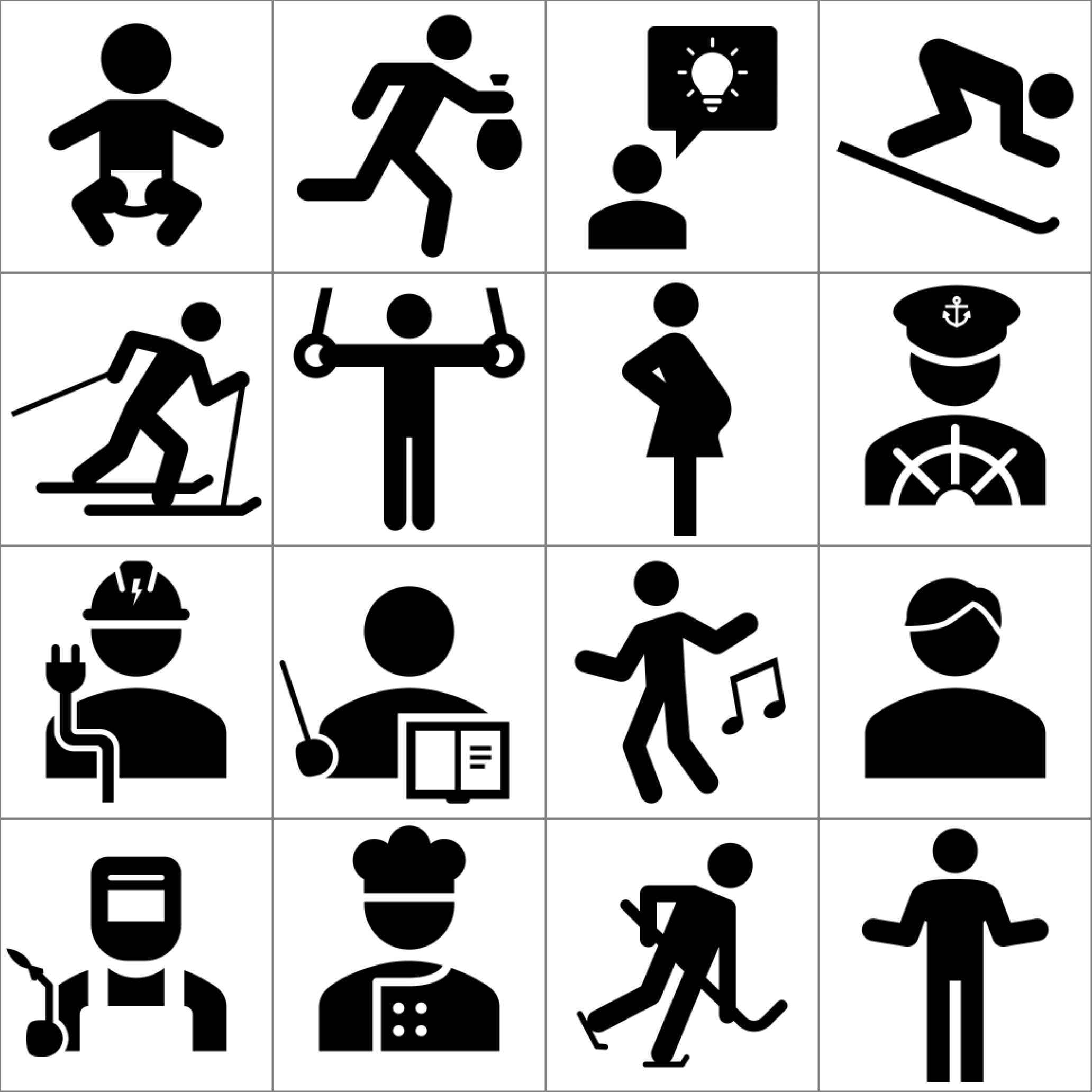}
  \small{(b)}
  \end{minipage}\\[-3mm]
\caption{(a)~Person photos and (b)~person icon images.\label{fig:person_sample}}
\vskip -3mm
\end{figure}

Since icons have no background in general, we need to prepare object images without background. Unfortunately, there is no large-scale image dataset that satisfies this condition. We, therefore, resort to MS-COCO~\cite{MSCOCO}, which is an image  dataset with pixel-level ground-truth for semantic segmentation. Figure~\ref{fig:COCO_sample} shows an image from MS-COCO and its pixel-level ground-truth for three objects, ``person'', ``dog'', and ``skateboard''.
Including those three classes, MS-COCO provides ground-truth for 80 object classes. \par
Figure~\ref{fig:obj_sample} shows examples of object images extracted by using the pixel-level ground-truth. 
After removing very small objects, we get 11,041 individual objects from 5,000 images of the MS-COCO. 
Those images were resized to be 256$\times$256 pixels including a white margin. 
Note that obtained object images often do not include the whole object. 
Thus, a part of an object is missed in most samples due to the occlusion in the original image. 
In addition, the object boundary is often neither smooth nor accurate.
Therefore, these object images are not perfect as the training samples for icon generation, although they are the best among the available datasets.

\begin{figure}[t] 
    \centering
    \includegraphics[width=0.48\textwidth]{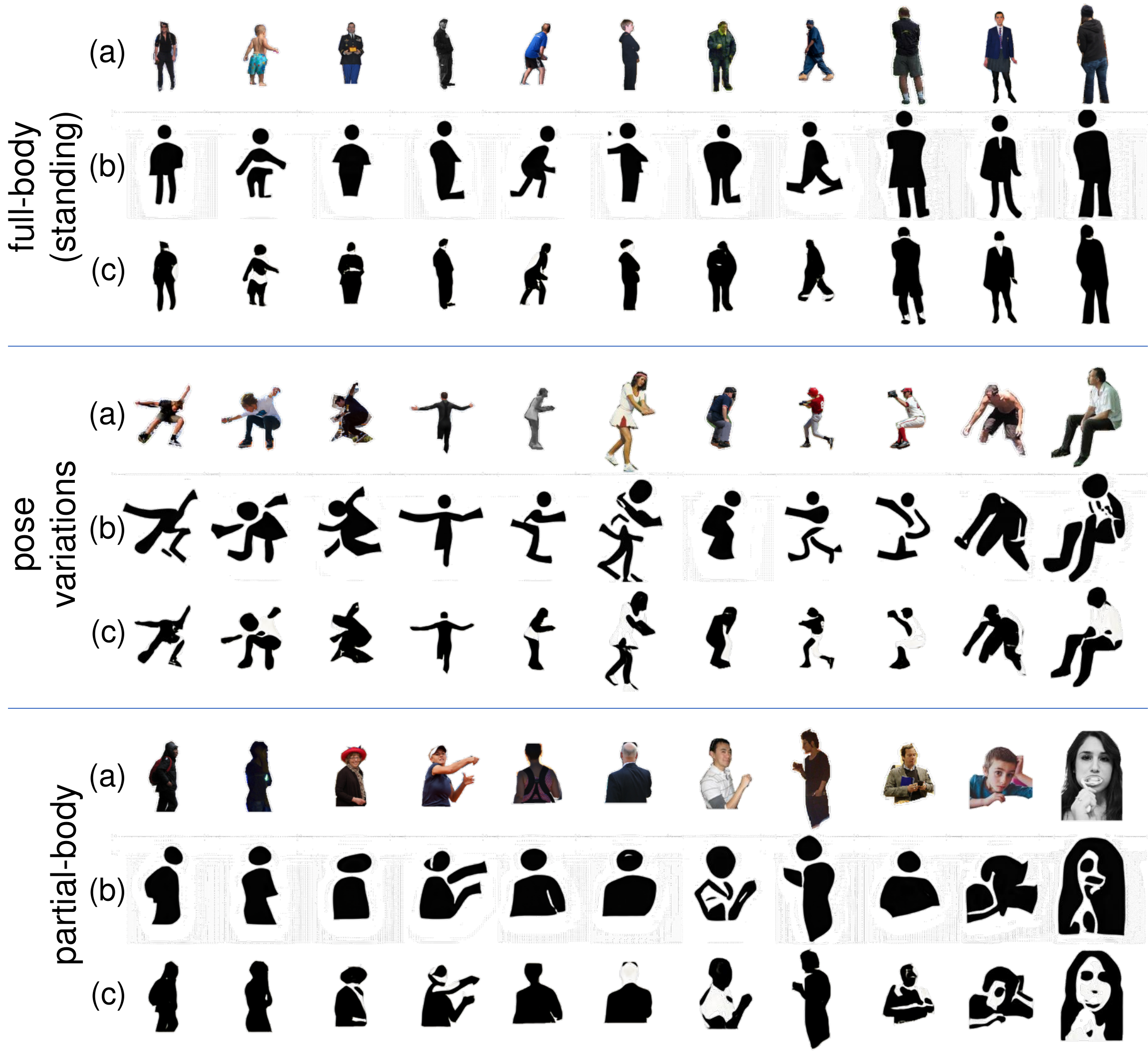}\\[-3mm]
    \caption{Iconified person photos by GANs trained with icons and photo images depicting persons. (a)~Original person photo. (b)~Iconified result by CycleGAN. (c)~Iconified result by UNIT.}
    \label{fig:person_result}
    \bigskip\bigskip
    \includegraphics[width=0.47\textwidth]{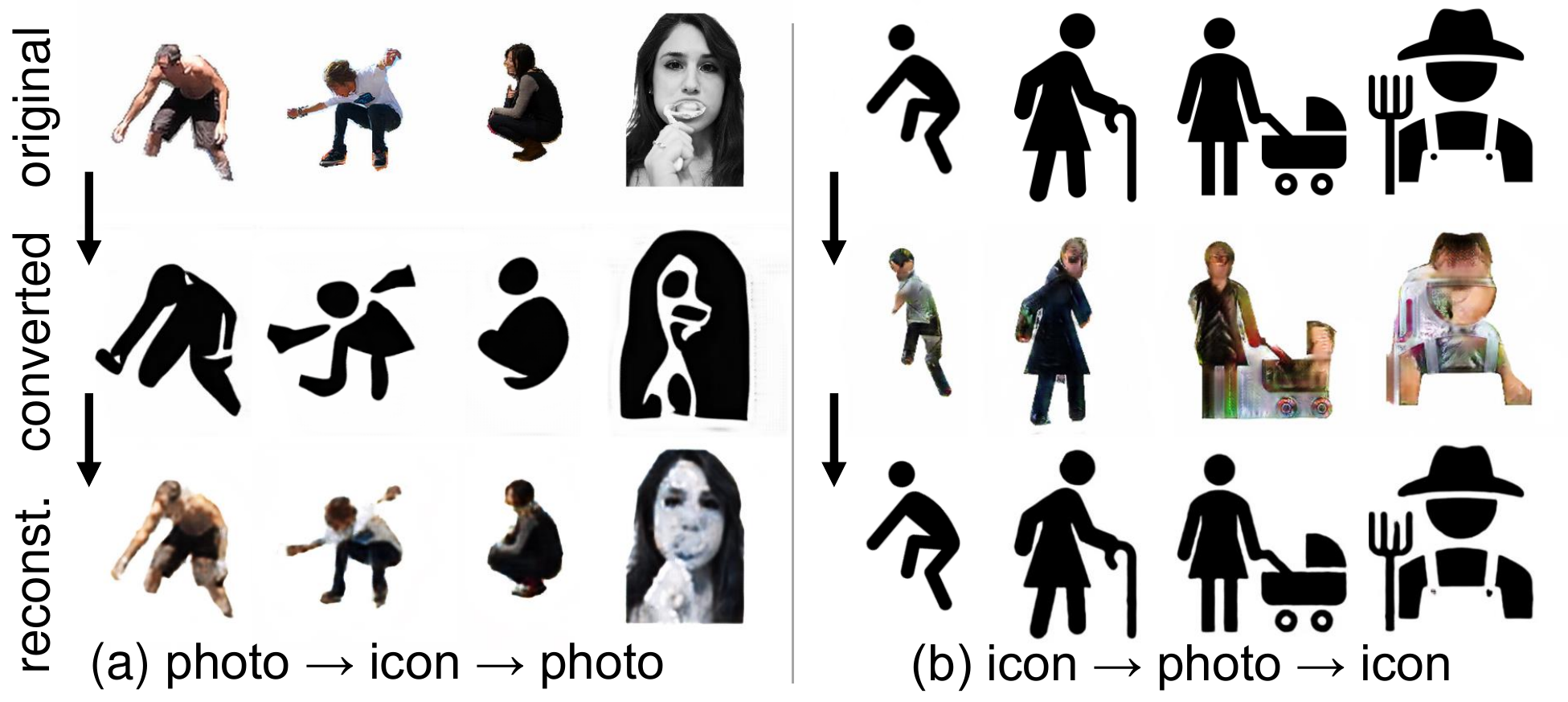}\\[-3mm]
    \caption{Reconstruction results by CycleGAN in two scenarios, (a) and (b).\label{fig:reconst}}
    \vskip -3mm
\end{figure}
\subsection{Icon image data}
As an icon image dataset, we used black-and-white icon images provided by Microsoft PowerPoint. Figure~\ref{fig:icon_sample} shows examples. Those icons are categorized into 26 classes and the total number of images is 883. Those images are resized to to be 256$\times$256 pixels including a white margin. As data augmentation during the training of GAN, they are translated, rotated, and scaled to increase their number up to 8,830.

\subsection{Logo image data as an alternative to icon images}
As an alternative to PowerPoint icons, we also examine logo images from LLD~\cite{LLD}. Logos and icons are different in their purpose and shape. For example, texts are often used in logos but not in icons. In addition, we can find more colorful images for logos than icons. However, they are still similar in their abstract design and therefore we also examine logo images.
Figure~\ref{fig:logo_sample} shows logo examples from LLD-logo. The 122,920 logo images in LLD-logo were collected from twitter profile images. In our experiment, we select  20,000 images randomly and resize them to be 256$\times$256 pixels (including a white margin) from their original 400$\times$400 pixels.
\section{Experimental results\label{sec:experiment}}
\subsection{Iconify human photos}
\label{sec:ex1}
As the first task, we train both GANs using only icons and photo images depicting persons. 
Figure~\ref{fig:person_sample} shows those training samples. By limiting the shape diversity in the training samples, we can observe the basic ability of GANs to iconify. 
In advance to training, we excluded person images which only capture a small part of a human body, such as hand and ear. 
Icon images showing multiple persons are also excluded. 
Finally, 1,440 icon images augmented from 72 icon images and 1,684 person photos are used as training samples for CycleGAN or UNIT in this experiment. 
\par

\begin{figure*}[t] 
    \centering
    \includegraphics[width=1\linewidth]{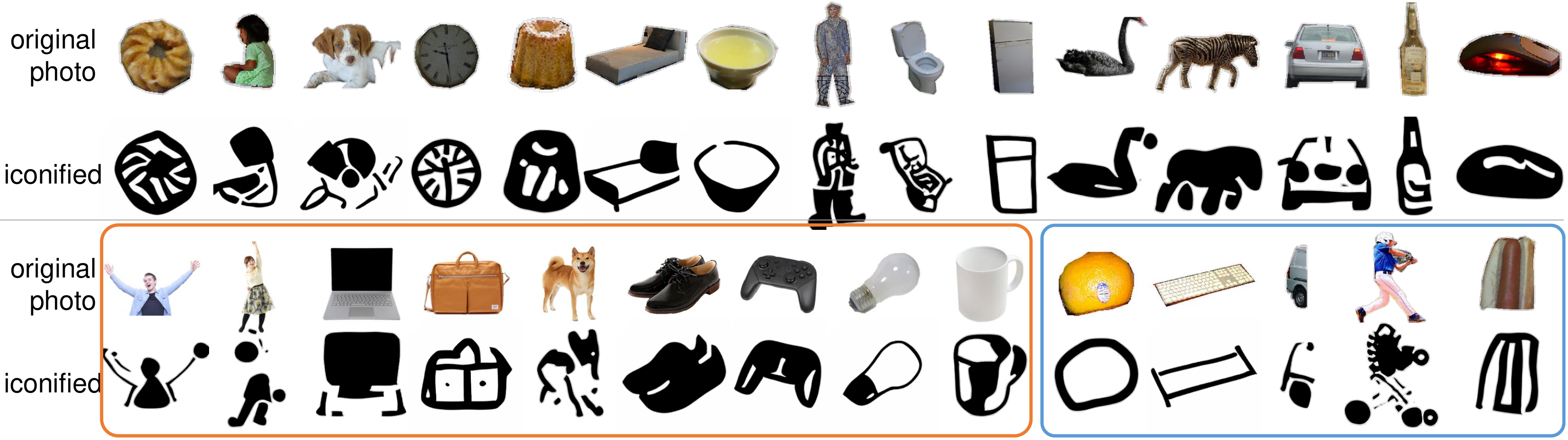}\\[-4mm]
    \caption{Iconified general image photos by CycleGAN trained with PowerPoint icon images. In the orange box, results of untrained samples are shown. In the blue box, typical failure results are shown. \label{fig:iconified-with-icon}}
    \bigskip\bigskip
    \includegraphics[width=\textwidth]{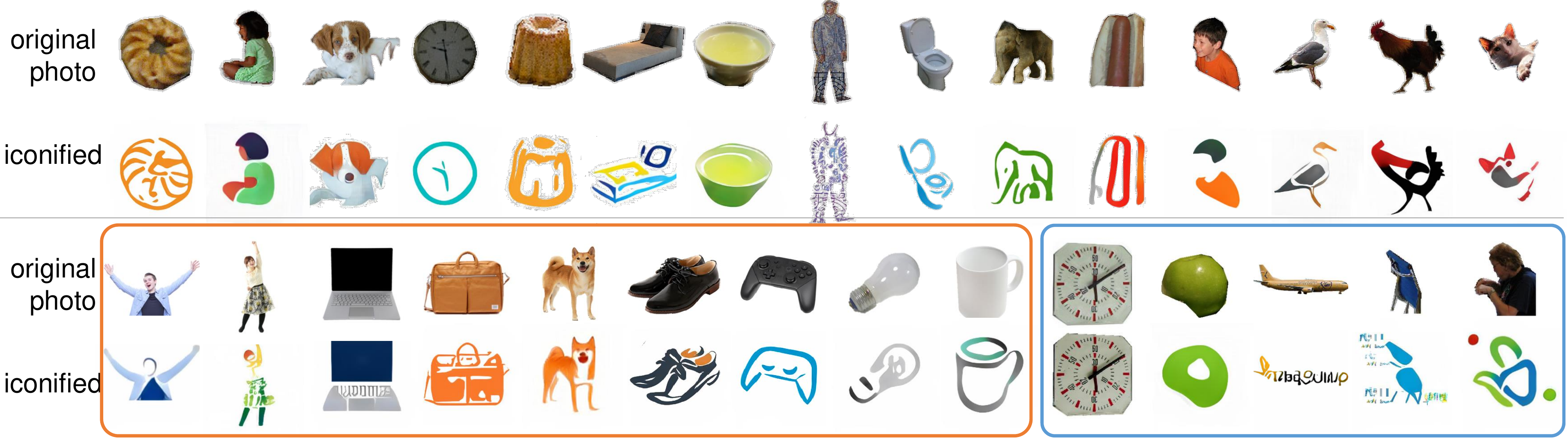}\\[-4mm]
    \caption{Iconified general image photos by CycleGAN trained with LLD-logo images. In the orange box, results for untrained samples are shown. In the blue box, typical failure results are shown.\label{fig:iconified-with-logo}}
    \vskip -2mm
\end{figure*}

Figure~\ref{fig:person_result} shows iconified person photos by CycleGAN and UNIT. These result images are the iconified results of the training samples. Since the number of images is very limited for this ``person-only'' experiment, it was not realistic to separate the images for training and testing. It should be noted that showing the results of the training samples is still reasonable. This is because, in our task, there is no ground-truth of the iconified result for each photo image; in other words, we do not use any ground-truth information during training. The results in the later sections contain the iconified results of the untrained samples.\par

From Figure~\ref{fig:person_result} we can see that both GANs successfully convert person photos into icon-like images; they are not just a binarization result but showing strong shape abstraction.
Especially, CycleGAN generates more abstract icon images with a circular head and a simplified body shape.  It is noteworthy that the head is often separated from the body part and it makes the generated images more icon-like. For facial images (in the bottom row), their iconified results are not natural. This is because we did not use icon images that show facial details during training.\par
Comparing to CycleGAN, the results by UNIT are less abstract (i.e., keeping the original shape of person photo) and therefore more similar to the binarization results. Since UNIT has a strong condition that the original photo and its iconified image share the same latent variable, it was difficult to realize strong shape abstraction. \par
Since CycleGAN has the cycle-consistency loss, it is possible to reconstruct the original photo image from its iconified versions.
Figure~\ref{fig:reconst}~(a) shows several reconstruction results. It is interesting to note that the original color image is still reconstructed from the black-and-white iconified result. It is also interesting to note that we can convert icon images to photo-like images by using the same CycleGAN model. The examples in Figure~\ref{fig:reconst}~(b) show the difficulty of this icon-to-photo scenario. However, the reconstructed icon images are almost the same as the original ones.
%
\subsection{Iconify general object photos with PowerPoint icons}
\label{sec:ex2}
As the second task, we use all photos from MS-COCO (Figure~\ref{fig:obj_sample}) and all icon images from PowerPoint (Figure~\ref{fig:icon_sample}) to train CycleGAN and generate the inconified results of general object photos. Since the first task reveals that CycleGAN has more abstraction ability than UNIT, we only use CycleGAN in this experiment. 
\par
This task is far more difficult than the previous; this is because CycleGAN needs to deal with not only the shape variations by the abstraction in icon images but also the shape variations by different object types (e.g., cars and balls). Moreover, the shape variations of object photo images are very severe due to the partial occlusions and non-accurate extractions, as noted in \ref{sec:obj_sample}.
\par
To deal with the huge variations, we used a simple coarse-to-fine strategy for training CycleGAN.
Specifically, we first train CycleGAN with the training samples resized to be 32$\times$32. Then, we fine-tune the CycleGAN with 64$\times$64, then 128$\times$128, and finally 256$\times$256. Similar coarse-to-fine strategies are used for other GANs, such as PGGAN\cite{PGGAN}, SinGAN\cite{SinGAN}, and DiscoGAN\cite{DiscoGAN}.
\par
Figure~\ref{fig:iconified-with-icon} shows the iconified results. The top row shows the results of the training samples (as noted \ref{sec:ex1}, showing the result of training samples is still reasonable since our framework is based on CycleGAN and there is no ground-truth). The results in the orange box of the bottom row show the results of  untrained samples (collected from copyright-free image sites). The iconified images show reasonable abstraction from the original photo images and it makes the iconified images different from binarization and edge extraction images. \par
Although the iconified images are promising to give a hint of icon design, the abstraction is not so strong as Figure~\ref{fig:person_result} of the first task. In addition, the iconified results are different from our ``standard'' icons. For example, the iconified doughnut and clock images in Figure~\ref{fig:iconified-with-icon} are different from the standard doughnut and clock icons in Figure~\ref{fig:icon_sample}, respectively. Since there is neither a common rule nor a strong trend in designing the standard icons of various objects, our iconified results show those differences.\par
The results in the blue box of Figure~\ref{fig:iconified-with-icon} are typical failure cases. From left to right, the first (orange) and second (keyboard) cases show too much abstraction. Since the original photo images are rather plain, the iconified results also become rough contour images. The third (car) case shows just a fragment of a car and the result cannot represent any car-like shape. The fourth (person) shows blob-like spurious noise, which are caused by insufficient training steps; in fact, in the early steps of CycleGAN training, we often find such failures.
\par
The last failure (hot dog) is an interesting but serious case. Although abstraction has been made appropriately, we cannot identify this iconified result as a hot dog. This case suggests that we need to be careful of the selection of the photo image for making its icon --- hot dog has its best appearance, shape, posture, and view angle for a legible icon. Non-legible iconified results occur in other objects by this reason. 
\subsection{Iconify general object photos with logos}
\label{sec:ex3}
Figure~\ref{fig:iconified-with-logo} shows the iconified results by CycleGAN trained with logo images from LLD\cite{LLD}. The top row shows the results of training samples (i.e., the object images from MS-COCO) and the orange box in the bottom row shows the results of the untrained samples. The photo images are converted like illustrations and therefore we can confirm CycleGAN can generate color icons. In some iconified results, the outline (i.e., edges) of the object is emphasized. \par
Comparing to the second task, it is also observed that the legibility of the icon images is greatly improved by color. For example, the hot dog icon in the top row shows better legibility than its black-and-white version in Figure~\ref{fig:iconified-with-icon}. Other iconified results also depict their original object more easily than black-and-white versions, even though the colors in the iconified images are not the same as the original object colors.
\par
In the blue box of Figure~\ref{fig:iconified-with-logo}, five typical failure cases are shown: from left to right, no significant change, too much abstraction, text-like icon, text-like spurious noise, and blob-like spurious noise.
The first case often occurs when the input photo shows a large object with no background part or a single-color object. The second occurs at fragmentary objects. The third occurs at flat objects; this is maybe due to many logo images from LLD contain a text part.  

\section{Conclusion and future work}
In this paper, we experimentally proved that the transformation of natural photos into icon images is possible by using generative adversarial networks (GAN). Especially, CycleGAN~\cite{CycleGAN} has a sufficient ``abstraction'' ability to generate icon-like images. For example, CycleGAN can generate person icons where each head is represented as a plain circle separated from the body part. From the qualitative evaluations, we can expect that the generated (i.e., iconified) images will give hints to design new icons for some object, although the iconified images sometimes show unnecessary artifacts or severe deformations.\par
As future work, it is better to conduct a subjective or objective evaluation of quality of the iconified images. Finding a larger icon dataset is also necessary to improve the quality. A more interesting task is the analysis of the trained GANs for understanding how the abstraction has been made; this will deepen our understanding about the strategy of professional graphic designers.

\begin{acks}
This work was supported by JSPS KAKENHI Grant Number \linebreak JP17H06100.
\end{acks}



\end{document}